\documentclass[12pt]{article}

\usepackage{graphicx}
\usepackage{epsfig}
\usepackage{amsfonts}
\begin{document} 

\title{On cosmological solutions in a spontaneously 
\\
broken gauge theory}
\author{{\large Yves Brihaye \footnote{yves.brihaye@umh.ac.be}}\\
\small{
Facult\'e des Sciences, Universit\'e de Mons-Hainaut,
B-7000 Mons, Belgium }\\
{ }\\
{\large Betti Hartmann \footnote{b.hartmann@iu-bremen.de}}\\
\small{
School of Engineering and Science, International University Bremen,
28725 Bremen, Germany
 }\\
{ }\\
{\large Eugen Radu\footnote{radu@thphys.nuim.ie}}\\
\small{
Department of  Mathematical Physics,
National University of Ireland Maynooth, Ireland}}

\date{\today}

\maketitle
\begin{abstract}
We consider solutions of the
Yang-Mills-Higgs system coupled to gravity in asymptotically
de Sitter spacetime. The basic features of two classes of solutions are discussed,
one of them corresponding to magnetic monopoles, the other one to sphalerons.
We find that although the total mass within the cosmological 
horizon of these configurations is finite,
their mass evaluated at timelike infinity generically diverges.
Also, no solutions exist in the absence of a Higgs potential.
 
\end{abstract}

%%%%%%%%%%%%%%%%%%%%%%%%%%%%%%%%%%%%%%%%%%%%%%%%%%%%%%%%%%%%%%%%%%%%%%%%%%%%%%
%%%%%%                     Introduction
%%%%%%%%%%%%%%%%%%%%%%%%%%%%%%%%%%%%%%%%%%%%%%%%%%%%%%%%%%%%%%%%%%%%%%%%%%%%%%
\noindent{\textbf{Introduction.--~}}
Some time ago it has been found that spontaneously broken
gauge theories admit classical, particle-like
solutions. The monopole \cite{'tHooft:1974qc}
and the sphaleron \cite{Dashen:1974ck} are the best known examples and  
physically the most relevant.
The magnetic monopoles 
inevitably arise in grand unification theories
and are stabilized by a quantum number of 
topological origin, corresponding to their magnetic charge.
Although the sphaleron solutions are unstable, they play an important
role in electroweak theory, fixing the energy
barrier separating topologically inequivalent vacua.

The effects of the gravitational self-interaction on magnetic
monopoles and sphalerons
have been addressed by many authors (see \cite{Volkov:1999cc} for a review).
Gravitating solutions exist up to
some maximal value of the coupling constant 
$\alpha$ of the theory 
(which is proportional to the ratio of the vector meson mass and Planck mass). 
Apart from the solutions with a regular origin, there are also nonabelian
``coloured'' black holes, parametrized by their event horizon radius.

However, most of the investigations in the literature 
have been carried out on the assumption that spacetime 
is asymptotically flat.
Less is known when the theory is modified to include a cosmological constant 
$\Lambda$ which greatly changes the asymptotic structure 
of spacetime \cite{Hawking}.
While
Einstein-Yang-Mills (EYM) configurations 
in asymptotic anti-de Sitter (AdS) space present a variety of
new qualitative features \cite{Winstanley:1998sn}, 
the solutions of a spontaneously
broken gauge theory in AdS are rather similar to the 
asymptotically flat counterparts \cite{Lugo:1999fm}, \cite{VanderBij:2001ah}.
Nontrivial solutions exist for any $\Lambda<0$;
as a new feature, one finds a complicated power decay of the fields
at infinity and a decrease of the 
maximal allowed vacuum  expectation value of the Higgs field.

For a positive cosmological constant, 
the natural ground state of the theory corresponds to
de Sitter (dS) spacetime.
This spacetime has gained a huge interest in theoretical physics recently for a
variety of reasons.
First of all, the observational evidence accumulated in the last
years \cite{data} is in favour of the idea that the
physical universe has an accelerated expansion. 
The most common explanation
is that the expansion is driven by a small positive vacuum energy
($i.e.$ a cosmological constant $\Lambda>0$).
Furthermore, dS spacetime plays a central
role in the theory of inflation.
Another motivation for studying dS spacetime is connected 
with the proposed holographic duality
between quantum gravity in dS spacetime and a conformal field theory
on the boundary of dS spacetime (see \cite{Klemm:2004mb} for a recent review
of this subject). 

In view of these developments, an examination of 
the classical solutions of gravitating 
fields in asymptotically  dS spacetimes seems appropriate.
The physical relevant case of a spontaneously  broken 
nonabelian gauge theory is particularly interesting, since it presents
particle like solutions with the same causal structure as dS spacetime.
Here we argue that the  features of these configurations
are rather different as compared to
the asymptotically flat of AdS counterparts.

%%%%%%%%%%%%%%%%%%%%%%%%%%%%%%%%%%%%%%%%%%%%%%%%%%%%%%%%%
\noindent{\textbf{The model.--~}}
%%%%%%%%%%%%%%%%%%%%%%%%%%%%%%%%%%%%%%%%%%%%%%%%%%%%%%%
We consider the action principle
\begin{eqnarray}
\label{lag0}
S=\int \sqrt{-g} d^4x \left( \frac{1}{16\pi G}
(R-2 \Lambda) +\mathcal{L} \right),
\end{eqnarray}
describing  Einstein gravity with a cosmological term coupled
to a Yang-Mills-Higgs (YMH) theory with compact gauge group 
$\mathcal{G}$ defined by the Lagrangian:
\begin{eqnarray}
\label{lagL} 
\mathcal{L} = -\frac{1}{4g^2} F_{\mu \nu} F^{\mu \nu}
-\frac{1}{2}(D_{\mu}\Phi)(D^{\mu}\Phi) - V(\Phi).
\end{eqnarray}
Here
$F_{\mu \nu}\equiv F_{\mu \nu}^a T_a=\partial_\mu 
A_\nu-\partial_\nu A_\mu+[A_\mu,A_\nu]$
is the gauge field strength, with the gauge field 
$A_\mu =A_\mu ^a T_a$, $T_a$ being the anti-Hermitian
gauge group generators and
$g$ the gauge coupling constant.
The Higgs field $\Phi$ is a vector in the representation space
of  $\mathcal{G}$ where the generators $T_a$ acts, with the covariant gauge derivative
$D_{\mu}\Phi=(\partial_\mu +A_\mu) \Phi$.

For simplicity we restrict ourselves to $\mathcal{G}$=SU(2), and a
double-well Higgs potential
$V(\phi)=\frac{1}{8}\lambda(\Phi^2-v^2)^2$. 
There are two
cases to be considered, leading to rather different
different types of solutions. The Higgs field can be chosen to be
either in the real triplet representation, in which case
$(T_a)_{ik}=-\epsilon_{aik}$ and we find monopole solutions, or in
the complex doublet representation with $(T_a)=\tau_a/2i$ ($\tau_a$
being the Pauli matrices), with sphaleron solutions.
 
We consider spherically symmetric configurations, with a line
element
\begin{equation}
\label{metric}
ds^{2}=\frac{dr^{2}}{N(r)}+r^2(d\theta^{2}+\sin^{2}\theta
d\varphi^{2})- \sigma^2(r)N(r)dt^{2}
\end{equation}
 where
\begin{equation}
N(r)=1-\frac{2m(r)}{r}- \frac{\Lambda }{3}r^2,
\end{equation}
%the function $m(r)$ being related to the local mass-energy density.
and a   gauge field ansatz  
\begin{equation}
\label{gauge} A=w(r)(-T_2 d \theta+T_1\sin \theta d\varphi)+T_3 \cos
\theta d \varphi.
\end{equation}
For the scalar field, we take $\Phi^a=\delta^3_k \phi(r)$ for a Higgs
triplet, and $\Phi^a=\xi^a \phi(r)$, with some constant spinor $\xi$,
for a Higgs field in the doublet representation.

The reduced EYMH action can be expressed as
\begin{equation}
S= \int dt~dr~\sigma \left[\frac{m'}{4 \pi G} -4 \pi \left(\frac{1}{g^2}
(Nw'^2+\frac{(1-w^2)^2}{2r^2} ) +\frac{1}{2}Nr^2 \phi'^2
+r^2V(\phi)+U(w,\phi) \right) \right],
\end{equation}
with
\begin{equation}
U(w,\phi)=w^2 \phi^2 {\rm ~~~resp.~~} U(w,\phi)=\frac{1}{4}(w+1)^2 \phi^2
\end{equation}
for the triplet respectively doublet Higgs. 
It is important to
notice that apart from the cosmological constant, the theory
contains three mass scales, the Planck mass $M_{Pl}=1/\sqrt{G}$, the
mass $M_W=gv$ of the YM field and the mass $M_H=\sqrt{\lambda}v$ of
the Higgs field.

Varying the reduced action one obtains the EYMH equations
\begin{eqnarray}
\label{e1} 
\nonumber
&&m' =4 \pi
G(\frac{1}{g^2}(\omega'^2N+\frac{(\omega^2-1)^2}{2r^2})
+\frac{1}{2}r^2N\phi'^2+U +V r^2 ),
\\
&&\sigma'=\frac{8\pi G
\sigma}{r}(\frac{1}{g^2}\omega'^2+\frac{1}{2}\phi'^2r^2),
~~~(N\sigma r^2 \phi')'=\sigma(\frac{\partial U}{\partial
\phi} +r^2\frac{d V}{d \phi}),
\\
\nonumber
&&(N\sigma \omega')'=\sigma \omega
\big(\frac{(\omega^2-1)}{r^2} + \frac{g^2}{2}\frac{\partial
U}{\partial w} \big),
\end{eqnarray}
where the prime indicates the derivative with respect to $r$.

Restricting to solutions with a regular origin,
we want the metric (\ref{metric}) to describe a nonsingular,
asymptotically de Sitter spacetime outside a cosmological horizon
located at $r=r_c>0$. Here $N(r_c)=0$ is only a coordinate
singularity where all curvature invariants are finite. A nonsingular
extension across this null surface can be found just as at the event
horizon of a black hole,
the Carter-Penrose conformal diagram being
qualitatively identical to
the de Sitter solution \cite{Volkov:1996qj}.

%%%%%%%%%%%%%%%%%%%%%%%%%%%%%%%%%%%%%%%%%%%%%%%%%%%%%%%%%%
\noindent{\textbf{Mass definition and asymptotic expansion.--~}}
%%%%%%%%%%%%%%%%%%%%%%%%%%%%%%%%%%%%%%%%%%%%%%%%%%%%%%%
The computation of the mass 
of asymptotically dS monopoles and sphalerons is
 a  difficult task due to the absence of spatial 
 infinity and the globally timelike Killing vector.
Also, these particle-like solutions typically strongly deform the
dS geometry inside the 
cosmological horizon.
Therefore, 
the perturbative approach measuring the energy of fluctuations
around the dS background proposed by Abbott and Deser may not be appropriate
in this case \cite{Abbott:1981ff}.
However, these  obstacles can be avoided
by using the prescription proposed in \cite{Balasubramanian:2001nb}
in which case the quasilocal
tensor of Brown and York (augmented by the AdS/CFT inspired counterterms
\cite{Balasubramanian:1999re}), 
is evaluated
on the Euclidean surfaces at future/past timelike infinity $\mathcal{I}^{\pm}$.
The conserved charge associated with the Killing vector $\partial/\partial t$
- now spacelike outside the cosmological horizon-
is interpreted as the conserved mass-energy $\mathfrak{M}$ \cite{Ghezelbash:2002ab}.
This allows also a discussion of the thermodynamics
of the asymptotically dS solutions outside the event horizon, 
the efficiency of this approach being demonstrated in a broad range of examples.
 
When applying this prescription to our case, we find that 
the asymptotic value of
the metric function $m(r)$ determines
the mass-energy of the monopole and sphaleron
solutions, $\mathfrak{M}=-\lim_{r \to \infty}m(r)$.

Following \cite{Gibbons:1977mu}, one may also define
a total mass 
$\mathbf{M}_c$ 
inside the cosmological
horizon.
This can be done by integrating the Killing identity
$\nabla^\mu\nabla_\nu K_\mu=R_{\nu\rho}K^\rho,$
 for the Killing field $K=\partial/\partial t$
on a spacelike hypersurface $\Sigma$ from the origin to
$r_c$ to get the Smarr-type formula
\begin{eqnarray}
\label{Smarr} 
\mathbf{M}_c \equiv \frac{1}{4 \pi G}\int \nabla_\mu K_\nu d\Sigma^{\mu \nu}=
%-\frac{\kappa_cA_c}{4\pi G}=
\frac{1}{4 \pi G}\int \Lambda K_\mu d\Sigma^\mu+
\int(2T_{\mu \nu}-Tg_{\mu \nu})K^\mu d\Sigma ^\nu.
\end{eqnarray}
%where $\kappa_c,~A_c$ are the cosmological horizon surface gravity and
%area, respectively.
It is natural to identify the left-hand side as the total mass within the cosmological
horizon. $\mathbf{M}_c$ can also be rewritten as $\mathbf{M}_c=- \kappa_cA_c/4\pi G=
-r_c^2\sigma(r_c) N'(r_c)/2G$,
where $\kappa_c,~A_c$ are the cosmological horizon surface gravity and
area, respectively.

In the vicinity of the origin, the solutions resemble the 
well known flat space configurations, with
 $w(0)=1$, $\phi(0)=0$ and $m(0)=0$.
The existence of a regular cosmological event horizon
at $r=r_c$ leads to the following conditions
\begin{eqnarray}
\label{cosm} 
\nonumber 
m(r_c)=\frac{r_c}{2}(1-\frac{\Lambda r_c^2}{3}),
~~
(N'\sigma\omega')\Big|_{r_c}=\sigma 
\left(\frac{\omega(\omega^2-1)}{r^2} + \frac{g^2}{2}\frac{\partial
U}{\partial w} \right)\bigg|_{r_c},
~~
N'\sigma r^2 \phi'\Big|_{r_c}=\sigma\left( \frac{\partial U}{\partial
\phi} +r^2\frac{d V}{d \phi}\right)\bigg|_{r_c}.
\end{eqnarray} 
 
%%%%%%%%%%%%%%%%%%%%%%%%%%%%%%%%%%%%%%%%%%%%%%%%%%%%%%%%%
%%%%  asymptotics la infinit
%%%%%%%%%%%%%%%%%%%%%%%%%%%%%%%%%%%%%%%%%%%%%%%%%%%%%%%%%
The boundary conditions at $r\to\infty$
are fixed by the requirements that the spacetime is asymptotically dS.
%, $i.e.$
%$g_{rr} \sim -\Lambda r^2/r+O(1/r^4)$, $g_{tt} \sim $.
When discussing the pure EYM system with $\Lambda>0$, there
are no restrictions on the asymptotic value of the gauge potential 
\cite{Volkov:1996qj}.
However, in the presence of a Higgs field, we find
that the gauge field should approach asymptotically
a fixed value $\omega_0$, which is zero for monopoles and $-1$ for sphalerons,
while the Higgs field reaches its vacuum expectation value.
This set of boundary conditions is shared also by asymptotically flat or AdS
configurations.  

However, the situation for $\Lambda>0$ is more subtle, since the 
cosmological constant enters in a nontrivial way the solutions' expression
as $r \to \infty$.
The analysis of the scalar field asymptotics
is standard; Strominger's mass bound 
 is $M_{S}^2= 3\Lambda/4$  \cite{Strominger:2001pn}
and separates the infinite energy solutions from solutions which may
present a finite mass (this would depend also on the gauge field
behaviour). For small enough values of the Higgs field mass,
$M_H<M_{S}$ the scalar field decays as
\begin{eqnarray}
\label{as1} 
\phi(r) \sim
v+c_1
r^{
-\frac{3}{2}
\left( 1+
\sqrt{1- M_H^2/M_{S}^2}
~\right)},
\end{eqnarray}
which assures a finite contribution
to the total mass-energy $\mathfrak{M}$ .

For a Higgs mass  exceeding Strominger's bound,
the scalar field behaves asymptotically as
%\begin{eqnarray}
%\label{as2} 
$\phi(r) \sim
v+c_2 r^{-3/2}\sin\left(\frac{3}{2}\sqrt{ M_H^2/M_S^2-1}~\log r+c_3\right)
$
%\end{eqnarray}
 which leads to a logarithmic divergence in the asymptotic expression
 of the mass function $m(r)$.
One may think that this bound may be circumvented 
by solutions with a vanishing Higgs potential.
However, 
by rewriting the Higgs field equation in the form
\begin{eqnarray}
\nonumber
\frac{1}{2}(N\sigma r^2 (\phi^2)')'&=&\sigma(N  r^2 \phi'^2+
\phi \frac{\partial U}{\partial \phi}
+\phi \frac{d V}{d \phi}),
\end{eqnarray}
and integrating it between the origin  and the 
cosmological horizon,
it can easily be proven that
no nontrivial solutions exist for $V(\phi)=0$ or for a convex potential.

A similar analysis reveals that
a positive cosmological constant sets  another mass bound for the
gauge sector, which is
 $M_{b}=\sqrt{\Lambda/12}$  for monopoles and $M_{b}=\sqrt{\Lambda/3}$
for sphalerons. 
Asymptotically dS solutions with a finite mass-energy exist for  
$M_W<M_b$, in which case the  expression of the gauge field as $r \to \infty$ is
\begin{eqnarray}
\label{as3} w(r) \sim
w_0+ c_4 r^{-\frac{1}{2}\left(1+\sqrt{1-M_W^2/M_b^2}\right)},
\end{eqnarray}
which contrasts with the exponential decay found in an asymptotically flat spacetime.
For $M_W> M_b$, the large $r$ behaviour of the gauge field is
%\begin{eqnarray}
% \label{as4}
$w(r) \sim w_0+ c_5 r^{-1/2} \sin
(\frac{1}{2}\sqrt{ M_W^2/M_b^2-1}\log r+c_6)$
%\end{eqnarray}
which leads to an infinite mass-energy $\mathfrak{M}$ of the configurations
(the constants $c_i$ which enter the above relations are free
parameters). 
The solutions with $M_H=M_{S}$,
 $M_W=M_b$ saturate these bounds and  lead also to infinite mass configurations.
Once we know the asymptotics of the matter fields,
the corresponding expression for the metric functions
 results
straightforwardly from the equations (\ref{e1}).

%%%%%%%%%%%%%%%%%%%%%%%%%%%%%%%%%%%%%%%%%%%%%%%%%%%%%%%%%
\noindent{\textbf{Numerical solutions.--~}}
%%%%%%%%%%%%%%%%%%%%%%%%%%%%%%%%%%%%%%%%%%%%%%%%%%%%%%% 
The solutions of the equations (\ref{e1}) are evaluated numerically.
With the boundary conditions discussed above, the procedure is to  
integrate separately between the origin and 
cosmological horizon and from the cosmological horizon 
to infinity, matching the solutions at $r=r_c$.
%The procedure is to integrate with the boundary conditions discussed above
%separately between the origin and 
%cosmological horizon and outside the cosmological horizon 
%to infinity and to match the solutions at $r=r_c$.

The usual  rescaling   
$r\to  g v r,$
$\phi\to\phi/v$  
 reveals  the existence of two  dimensionless  parameters $\alpha$ and $\beta$, 
 expressible  through the mass
ratios
$\alpha=M_{W}/M_{Pl},~
 \beta=M_{H}/M_{W}$.
The third parameter of the system is the rescaled cosmological constant 
$ \Lambda\to \Lambda G/g^2v^2$. 
The configurations with $\alpha=0$ correspond to
monopoles and sphalerons in a fixed dS background, and contain already the
basic features of the theory.

The equations of motion (\ref{e1}) were solved
varying $\Lambda$ for a range of values of the coupling
parameter $\alpha$ 
and several values of $\beta$. 
While a negative cosmological constant exerts an additional pressure on solitons, 
causing their
typical radius to become thinner \cite{Lugo:1999fm, VanderBij:2001ah},
a positive $\Lambda$ has the opposite effect, causing the soliton 
radius to expand beyond the value it would have
in asymptotically flat space.
Also, as $\alpha$ increases, the cosmological horizon shrinks in size.
The dS solitons are generally not confined inside the cosmological horizon,
with all variables and their first derivatives extending smoothly
through the cosmological horizon.
The profiles of typical solutions are presented in Figure 1.

 When $\Lambda$ is increased from zero, while keeping $\alpha,~\beta$ fixed,
a branch of dS solutions emerges from the corresponding asymptotically
flat
configurations.  This branch ends at
a maximal value 
%%%%%%%%%%%%%%%%%%%%%%%%%%%%%%%%%%%%%%%%%%%%%%%%%%%%%%%%%%%%%%%%%%%%%%%%%%
\newpage
\setlength{\unitlength}{1cm}
\begin{picture}(6,8)
\centering
\put(1.6,0){\epsfig{file=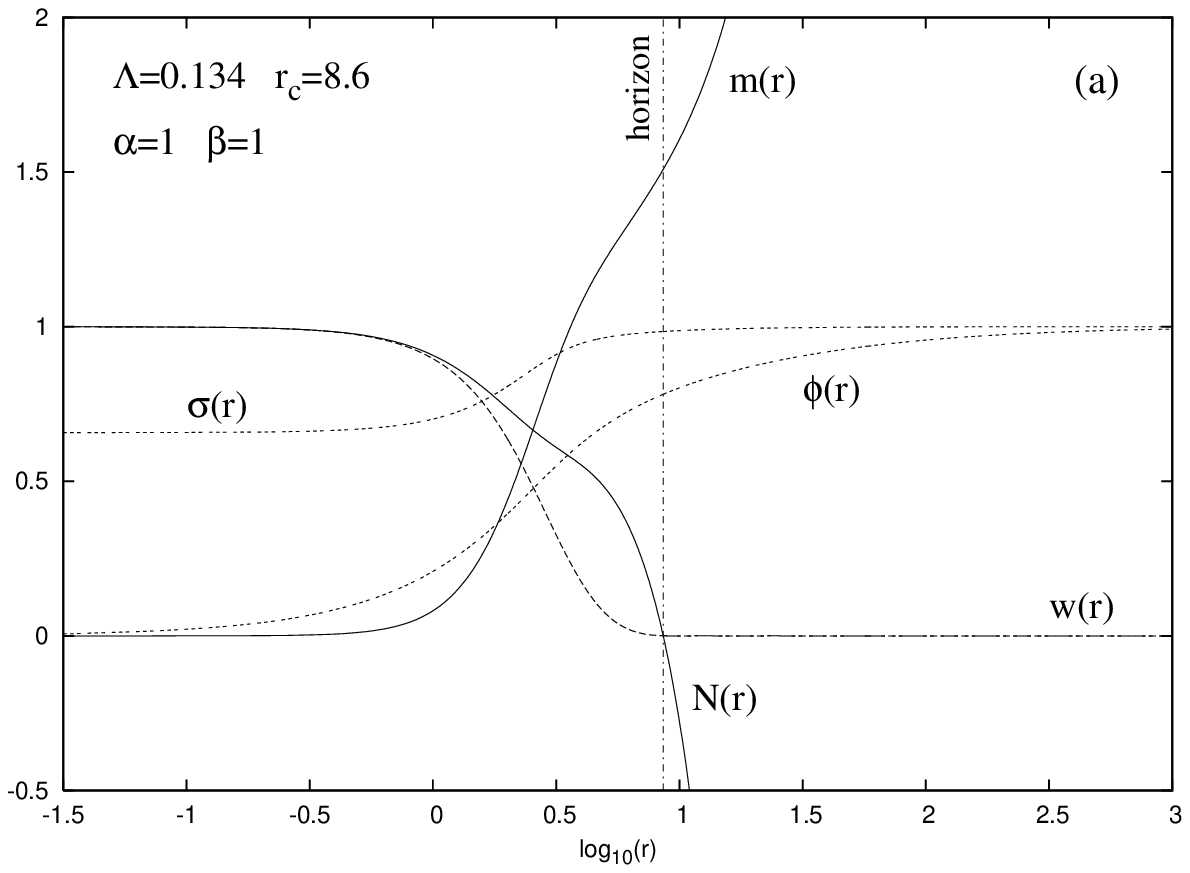,width=12cm}}
\end{picture}
\begin{center}
\end{center}
\begin{picture}(10,7.7)
\centering
\put(2.26,0){\epsfig{file=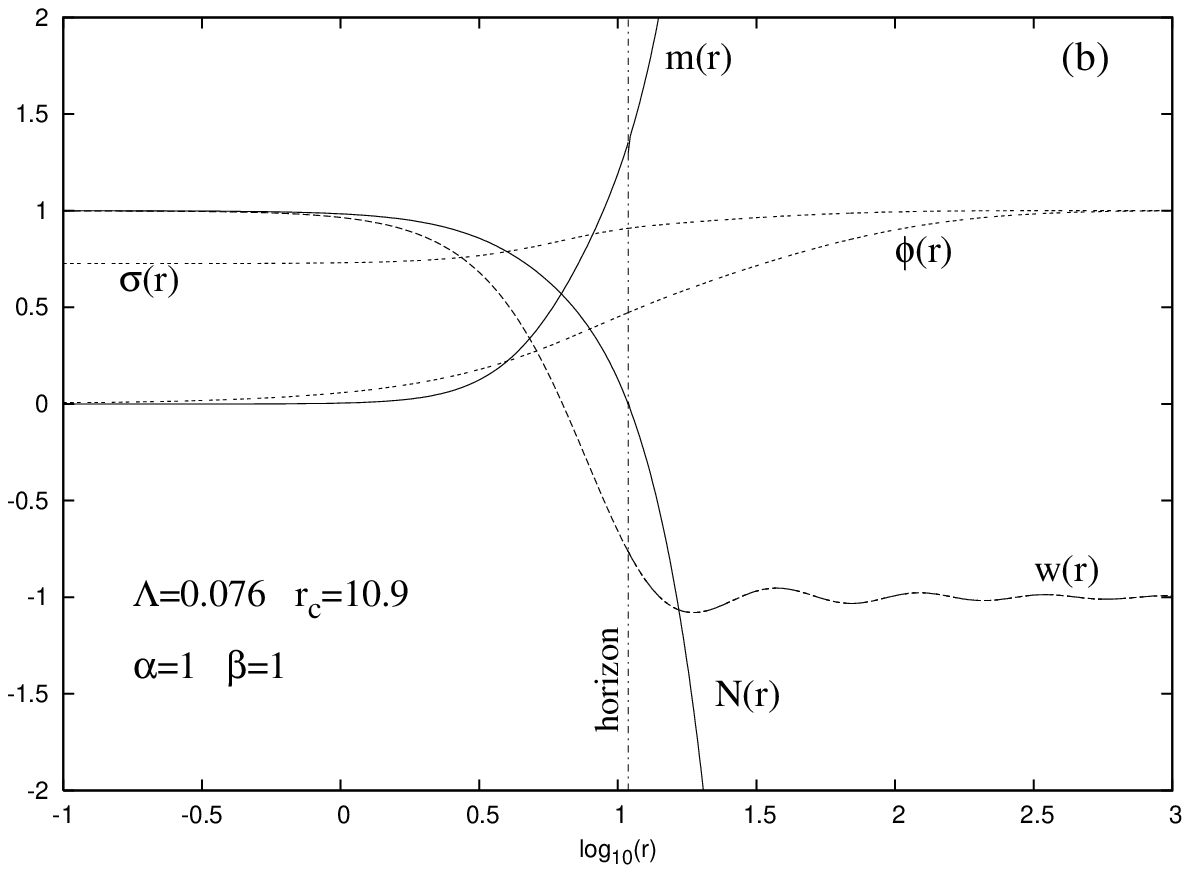,width=12cm}}
\end{picture}
\\
{\small {\bf Figure 1.}  Typical asymptotically de Sitter monopole (figure 1a) and 
sphaleron solutions (figure 1b). }
\\
\\
$\Lambda_{max}$. A second branch of  solutions  always  appears at $\Lambda_{max}$,
 extending  backwards in $\Lambda$ to a zero
value  of the cosmological constant (for monopoles) or to some  small $\Lambda_{c}\neq 0$
for sphalerons.
In this limit,  the trivial solution
 $\phi(r)=0,~w(r)=1$ is approached.
The value of $\Lambda_{max}$ depends on the parameters $\alpha$, 
$\beta$;
for example, for solutions with 
 $\beta=0.1$ in a  fixed dS background,
we find  $\Lambda_{max}\approx 0.069$  for monopoles
and   $\Lambda_{max}\approx 0.0506$  for sphalerons.
The value of $\Lambda_{max}$ is only slightly affected by changing $\alpha$,
$e.g.$  for solutions with $\alpha=1,~\beta=0.1$
we find $\Lambda_{max}\approx 0.067$ for monopoles and 
$\Lambda_{max}\approx 0.0505$ for sphalerons.

 %%%%%%%%%%%%%%%%%%%%%%%%%%%%%%%%%%%%%%%%%%%%%%%%%%%%%%%%%%%%%%%%%%%%%%%%%%
\newpage
\setlength{\unitlength}{1cm}
\begin{picture}(6,8)
\centering
\put(1.6,0){\epsfig{file=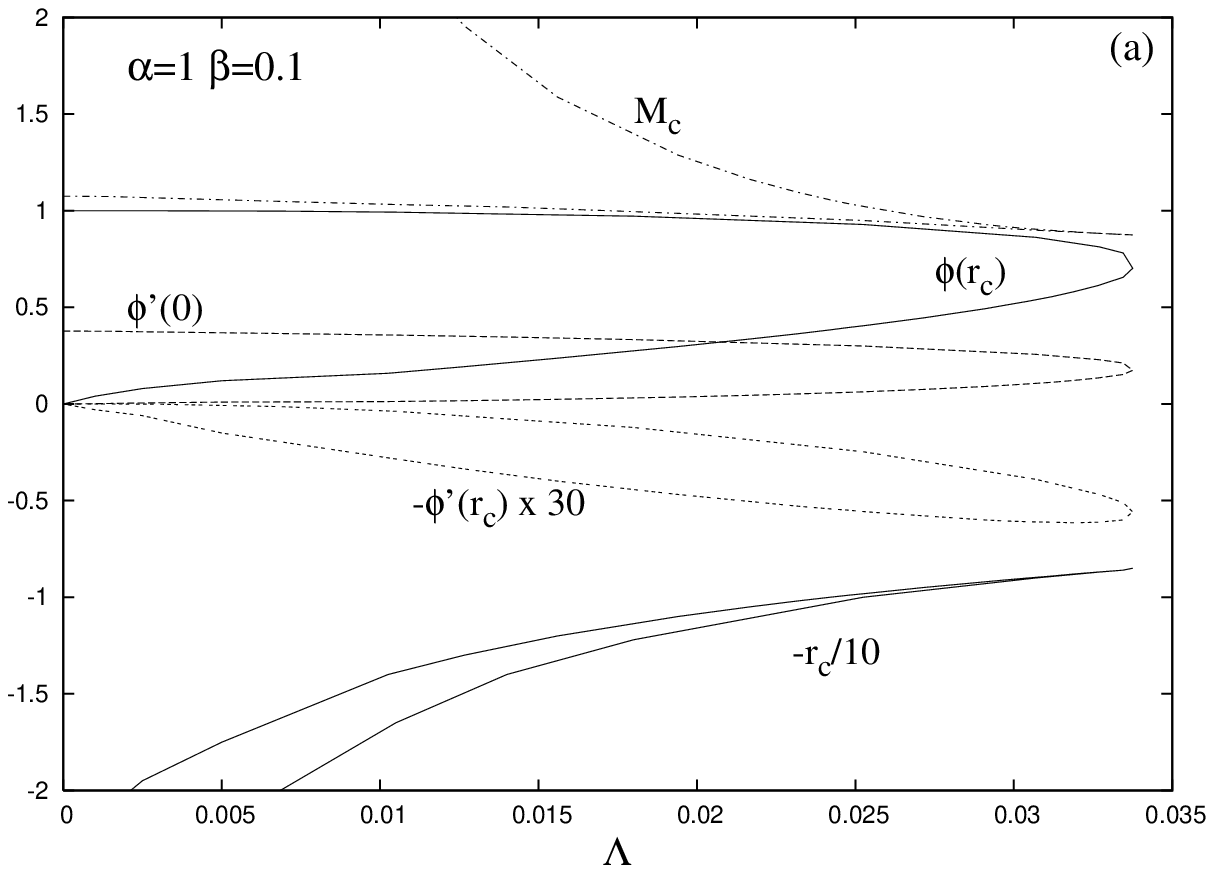,width=12cm}}
\end{picture}
\begin{center}
\end{center}
\begin{picture}(10,7.7)
\centering
\put(2.2,0){\epsfig{file=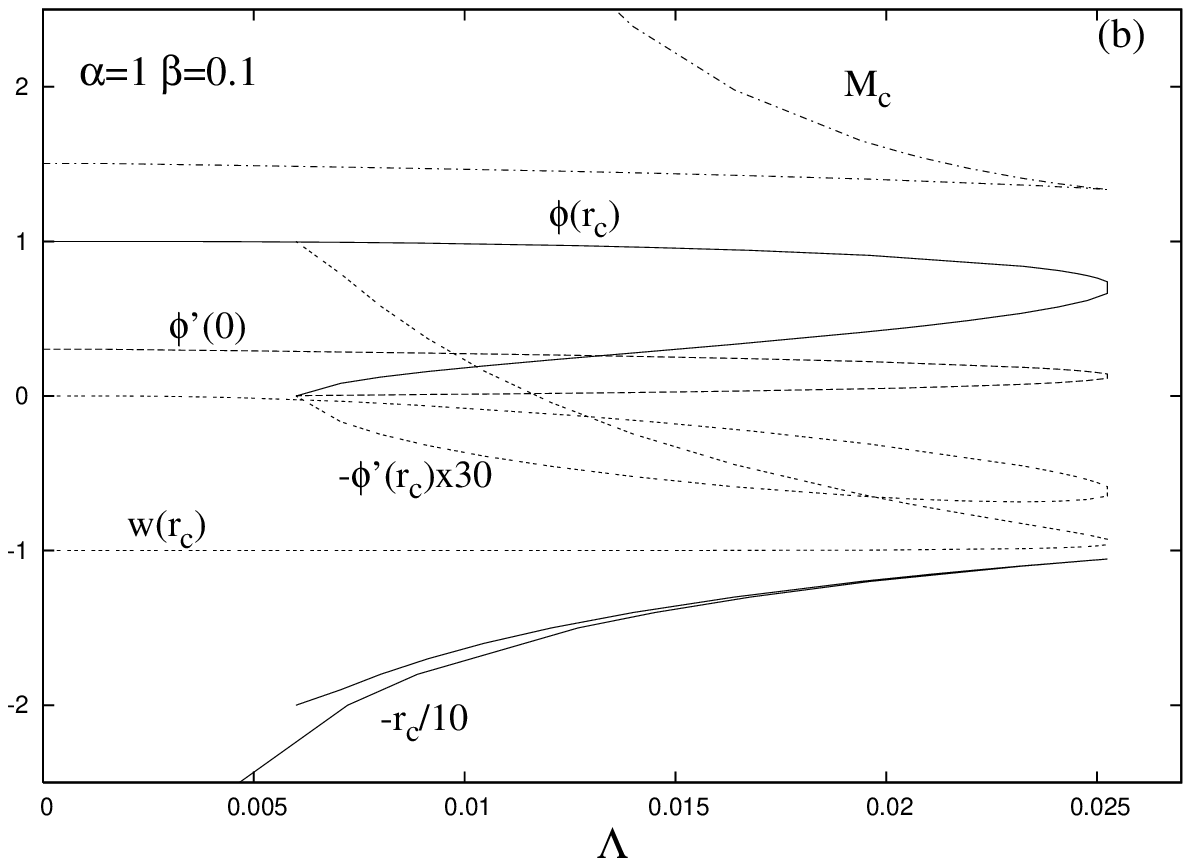,width=12cm}}
\end{picture}
\\
{\small {\bf Figure 2.}  The dependence
of solution properties on the value of the cosmological constant
is plotted for monopoles (figure 2a) and 
sphalerons (figure 2b). }
\\
\\
These statements are illustrated
 in Figure 2 where some numerical data is plotted as function
 of $\Lambda$ for the two branches, respectively for dS monopoles
 and dS sphalerons. 
The figures are obtained for 
$\alpha=1,~\beta=0.1$
but they remain qualitatively the same for all gravitating solutions
we considered.
The maximal values of  $\Lambda$ are always below the critical values found for solutions
in fixed dS background, and
as a result the mass of our solutions measured at timelike infinity
always diverges, although the mass $\mathbf{M}_c$ 
within the cosmological
horizon stays finite (see also Figure 1).
The existence of other disconnected branches of solutions
for $\Lambda>\Lambda_{max}$ appears unlikely.
Note that the EYM theory also presents solutions with dS asymptotics only for 
values of the cosmological constant up to some $\Lambda_{max}<3/4$ \cite{Volkov:1996qj}. 

For a given value of $\Lambda<\Lambda_{max}$, we notice  the existence
of  a maximal 
value of $\alpha$, which depends on $\beta$.
The behavior of solutions as $\alpha\to \alpha_{max}$ is similar to the 
asymptotically flat case.
The gravitating monopoles separate in this limit 
into an interior region with a smooth origin and a nontrivial YM field,
and an exterior extremal Reissner-Nordstr\"om-dS solution with $w=0$.
 Different from the monopole case, the sphaleron solutions
may be continued all the way back to $\alpha=0$,
where we end up with a cosmological
EYM solution.
 
%%%%%%%%%%%%%%%%%%%%%%%%%%%%%%%%%%%%%%%%%%%%%%%%%%%%%%%%%
\noindent{\textbf{Conclusions.--~}}
%%%%%%%%%%%%%%%%%%%%%%%%%%%%%%%%%%%%%%%%%%%%%%%%%%%%%%%
In this letter we discussed the basic properties of 
the monopole and sphaleron solutions in an asymptotically dS spacetime.
Contrary to the naive expectation
that a small $\Lambda$ will 
not affect the properties of the configurations drastically,
we find that the mass of dS solutions evaluated at timelike
infinity by using the quasilocal
tensor of Brown and York diverges
(although the mass within the cosmological horizon stays finite).
%Moreover, a positive cosmological constant
%sets an upper bound for the masses of  
%the vector boson and the Higgs field.
 These features are shared also by the 
 black hole counterparts of the soliton solutions,
 which can be constructed by using the same techniques.

A divergent ADM mass has been found also for solutions of some theories
in asymptotically AdS spacetime.
However, in some cases it is still  possible to obtain a finite mass by
allowing the regularizing counterterms to depend not only on
the boundary metric  but
also on the matter fields on the boundary \cite{Hollands:2005wt}.
It would be interesting to generalize this method to the dS case 
and to assign a finite mass (evaluated outside the cosmological horizon)
to the solutions of a spontaneously broken gauge theory.

We believe that this may lead to further understanding
of the rich structure of a field theory in dS space
as well as profound implications to the evolution of the 
early universe.

An extensive analysis of the solutions with variation of the parameters of the theory,
as well as black hole configurations, will be presented in a separate publication.
\\
\\
\\
{\bf Acknowledgements} 
YB is grateful to the
Belgian FNRS for financial support.
The work of ER is carried out
in the framework of Enterprise--Ireland Basic Science Research
Project
SC/2003/390 of Enterprise-Ireland.

%%%%%%%%%%%%%%%%%%%%%%%%%%%%%%%%%%%%%%%%%%%%%%%%%%%%%%%%%%%%%%%%%%%%%%%%%%%%%%

%%%%%%%%%%%%%%%%%%%%%%%%%%%%%%%%%%%%%%%%%%%%%%%%%%%%%%%%%%%%%%%%%%%%%%%%%%%%%%

\end{document}